\documentclass[aps, prl
, twocolumn
, nofootinbib
, superscriptaddress
]{revtex4-2}

\usepackage{graphicx}
\usepackage{xcolor}
\usepackage[caption=false]{subfig}
\usepackage{mathrsfs,mathtools}
\usepackage{physics,amssymb}
\usepackage{siunitx}
\usepackage{bm}
\usepackage{braket}
\usepackage{cases}
\usepackage{comment}
\usepackage{soul}
\usepackage{cancel}
\usepackage[utf8]{inputenc}
\usepackage{url}
\usepackage{multirow}
\usepackage{xspace}
\usepackage{acronym}
\usepackage[colorlinks=true
,urlcolor=DARKBLUE
,anchorcolor=DARKBLUE
,citecolor=DARKBLUE
,filecolor=DARKBLUE
,linkcolor=DARKBLUE
,menucolor=DARKBLUE
,linktocpage=true
,pdfproducer=medialab
,pdfa=true
]{hyperref}

\newcommand{\ee}{\mathrm{e}}
\newcommand{\Mpl}{M_\mathrm{Pl}}
\newcommand{\ns}{n_{\mathrm{s}}}
\newcommand{\cs}{c_{\mathrm{s}}}
\newcommand{\GW}{\mathrm{GW}}
\newcommand{\QCD}{\mathrm{QCD}}
\newcommand{\RD}{\mathrm{RD}}
\newcommand{\PG}{\mathrm{PG}}
\newcommand{\umin}{\mathrm{min}}
\newcommand{\umax}{\mathrm{max}}
\newcommand{\yr}{\mathrm{yr}}

\newcommand{\uc}{\mathrm{c}}

\newcommand{\uf}{\mathrm{f}}

\newcommand{\calH}{\mathcal{H}}

\newcommand{\bfk}{\mathbf{k}}

\newcommand{\calP}{\mathcal{P}}

\newcommand{\calS}{\mathcal{S}}

\newcommand{\beae}[1]{\begin{equation}\begin{aligned} #1 \end{aligned}\end{equation}}
\newcommand{\bege}[1]{\begin{equation}\begin{gathered} #1 \end{gathered}\end{equation}}
\newcommand{\bae}[1]{\begin{align} #1 \end{align}}

\newcommand{\bfe}[4]{
\begin{figure} 
	\centering
	\includegraphics[#1]{#2}
	\caption{#3}
	\label{#4}
\end{figure}}
\newcommand{\bme}[1]{\begin{multline} #1 \end{multline}}

\definecolor{MONZA}{HTML}{CF000F}
\definecolor{DARKBLUE}{HTML}{00008b}
\definecolor{DARKMAGENTA}{HTML}{8b008b}

\begin{document}
\title{Translating nano-Hertz gravitational wave background into primordial perturbations taking account of the cosmological QCD phase transition}
\date{\today}

\author{Katsuya T. Abe}
\email{kabe@chiba-u.jp}
\affiliation{Center for Frontier Science, Chiba University, 1-33 Yayoi-cho, Inage-ku, Chiba 263-8522, Japan}
\author{Yuichiro Tada}
\email{tada.yuichiro.y8@f.mail.nagoya-u.ac.jp}
\affiliation{Institute for Advanced Research, Nagoya University,
Furo-cho Chikusa-ku, 
Nagoya 464-8601, Japan}
\affiliation{Department of Physics, Nagoya University, 
Furo-cho Chikusa-ku,
Nagoya 464-8602, Japan}

\begin{abstract}
The evidence of the nano-Hertz stochastic \ac{GW} background is reported by multiple pulsar timing array collaborations. While a prominent candidate of the origin is astrophysical from supermassive black hole binaries, alternative models involving GWs induced by primordial curvature perturbations can explain the inferred GW spectrum. Serendipitously, the nano-Hertz range coincides with the Hubble scale during the cosmological \ac{QCD} phase transition. The influence of the QCD phase transition can modify the spectrum of induced GWs within the nano-Hertz frequency range, necessitating careful analysis. We estimate GWs induced by power-law power spectra of primordial curvature perturbations taking account of the QCD phase transition. Then we translate the implication of the NANOGrav data into the constraint on the power spectrum of the primordial curvature perturbation, 
which suggests one would underestimate the amplitude by about $25\%$ and the spectral index by up to $10\%$ if neglecting the \ac{QCD} effect.
\end{abstract}

\maketitle

\acrodef{PBH}{primordial black hole}
\acrodef{GW}{gravitational wave}
\acrodef{QCD}{quantum chromodynamics}
\acrodef{CMB}{cosmic microwave background}

\section{Introduction}

The evidence of the stochastic \ac{GW} background in the nano-Hertz range is reported by the NANOGrav~\cite{NANOGrav:2023gor}, European Pulsar Timing Array~\cite{Antoniadis:2023ott}, Parkes Pulsar Timing Array~\cite{Reardon:2023gzh}, and Chinese Pulsar Timing Array~\cite{Xu:2023wog}.
The inferred spectrum is consistent with the astrophysical expectation from supermassive black hole binaries, but it can be also explained by some primordial origin~\cite{NANOGrav:2023hvm,Antoniadis:2023zhi} represented by the induced \ac{GW} due to large primordial curvature perturbations~\cite{tomita1967non,Matarrese:1992rp,Matarrese:1993zf,Matarrese:1997ay,Carbone:2004iv,Ananda:2006af,Baumann:2007zm} (see, e.g., Ref.~\cite{Chen:2019xse} for the induced-\ac{GW} interpretation of the NANOGrav 11-yr data).
In particular, \acp{GW} with the observed amplitude and frequency range would correspond to sizable enough perturbations which can cause \acp{PBH} of the stellar mass~\cite{Saito:2008jc,Bugaev:2009zh,2010PThPh.123..867S,Bugaev:2010bb,Inomata:2016rbd} (see also the recent review article~\cite{Domenech:2021ztg}).
Such stellar mass \acp{PBH} can explain some fraction of black hole binaries found by merger \acp{GW} in the LIGO--Virgo--KAGRA collaboration~\cite{Bird:2016dcv,Clesse:2016vqa,Sasaki:2016jop,Sasaki:2018dmp}.
Large primordial perturbations themselves are viewed as important information on the detailed mechanism of cosmic inflation.

\begin{figure*}
    \centering
    \begin{tabular}{c}
        \begin{minipage}{0.5\hsize}
        \centering
        \includegraphics[width=0.8\hsize]{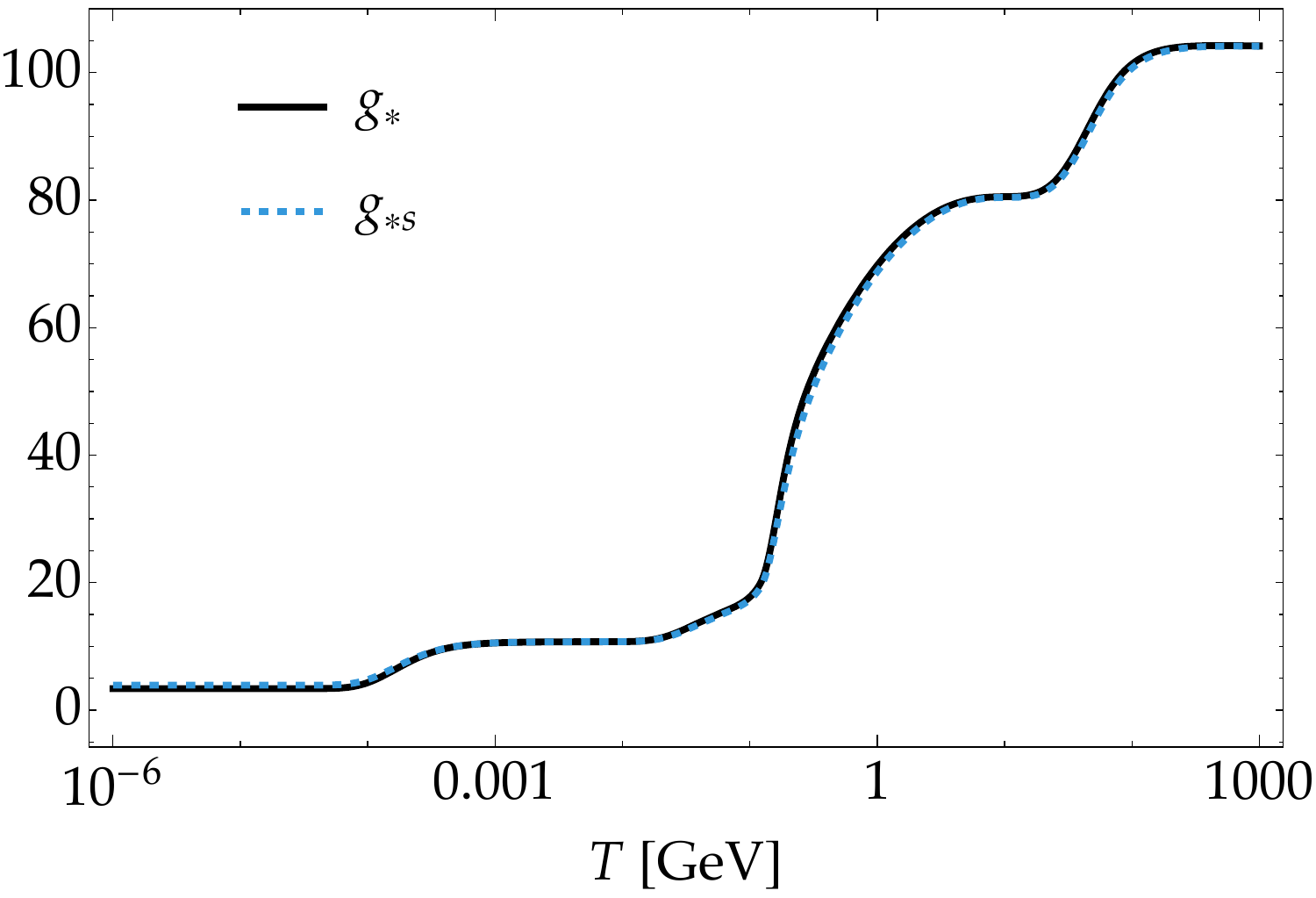}
        \end{minipage}
        \begin{minipage}{0.5\hsize}
        \centering
        \includegraphics[width=0.8\hsize]{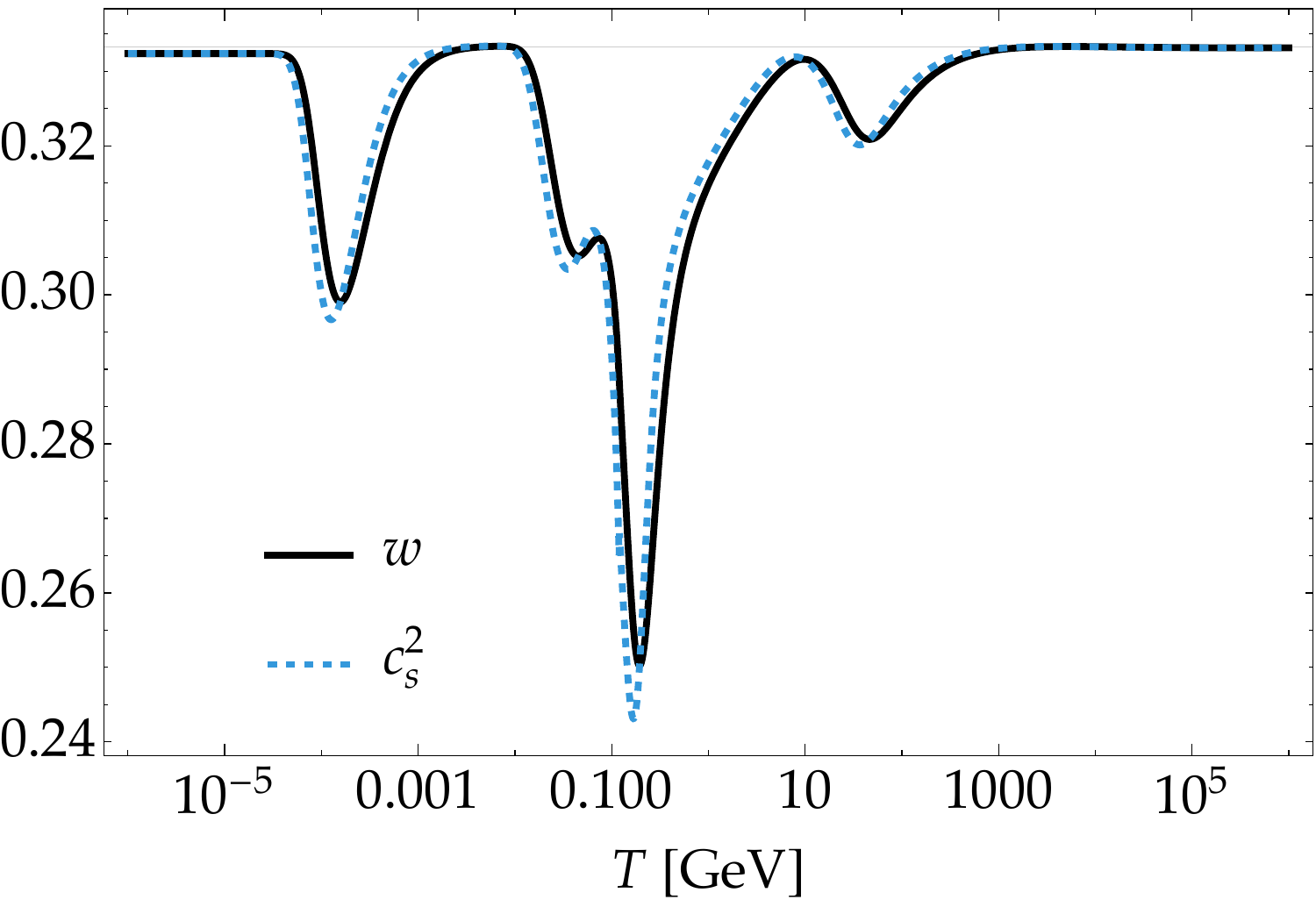}
        \end{minipage}
    \end{tabular}
    \caption{\emph{Left}: the temperature dependence of the effective degrees of freedom for energy density, $g_*$ (black), and entropy density, $g_{*s}$ (light-blue dashed). We adopt the fitting formula given in Appendix~C of Ref.~\cite{Saikawa:2018rcs} throughout this paper. \emph{Right}: corresponding equation-of-state parameter $w$ (black) and the sound speed squared $\cs^2$ (light-blue dashed)~\eqref{eq: w and cs2}. The horizontal line $w=\cs^2=1/3$ is the value of the exact radiation fluid.}
    \label{fig: gs and wcs}
\end{figure*}

\begin{figure}
    \centering
    \includegraphics[width=\hsize]{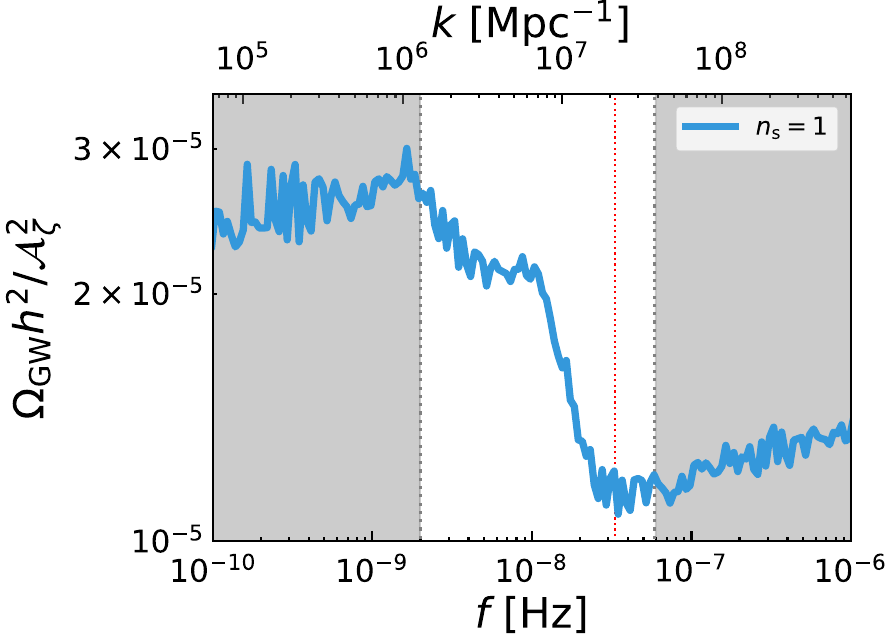}
    \caption{The spectral shape of the current energy density of GWs induced by the scale-invariant spectrum (light blue), whose amplitude is normalized by $A_\zeta^2$.
    The GW frequency $f$ is related with its wavenumber $k$ by $f=k/(2\pi)$. The noisy feature on the induced spectrum is merely caused by the numerical error. 
    The unshaded region shows the NANOGrav's sensitivity range $f=2\text{--}\SI{59}{nHz}$ and the red dotted line indicates the pivot scale $f_{\si{yr^{-1}}}=\SI{1}{yr^{-1}}\simeq\SI{31.7}{nHz}$ of the analysis.}
    \label{fig: OmegaGW}
\end{figure}

Serendipitously, the nano-Hertz range coincides with the Hubble scale during the cosmological \ac{QCD} phase transition.
There, the equation-of-state parameter $w=p/\rho$ and the sound speed (squared) $\cs^2=\pdv*{p}{\rho}$, where $\rho$ and $p$ are energy density and pressure, slightly reduce from the exact radiation value, $1/3$, (see Fig.~\ref{fig: gs and wcs}) and hence the compaction of the density perturbation and also the dilution of the induced \ac{GW} are affected~\cite{Abe:2020sqb}.
In fact, the resultant \ac{GW} spectrum shows a sharp drop in this range even if the input primordial curvature perturbation is exactly scale-invariant (see Fig.~\ref{fig: OmegaGW}). Hence, a naive estimate without the \ac{QCD} effect can miss the true implication.
In this \emph{Letter}, given input (compact)  power-law power spectra of curvature perturbations, we numerically calculate the resultant spectra of the induced \acp{GW} with the \ac{QCD} effect in the nano-Hertz range and derive fitting formulae for their amplitude and scale dependence with respect to the input parameters.
Making use of these formulae, specifically, the implication of the NANOGrav data (see Fig.~\ref{fig: NANOGrav const}) is translated into the constraint on the power spectrum of the primordial curvature perturbation.\footnote{See Refs.~\cite{Inomata:2023zup,Cai:2023dls,Wang:2023ost,Liu:2023ymk} for the implication on the \ac{PBH} and induced \ac{GW} (without the \ac{QCD} effect) of the latest NANOGrav data. See also Ref.~\cite{Franciolini:2023wjm} for a discussion of the QCD effect (only on $w$) on the primordial GW in general contexts in nano-Hertz frequency ranges.}
Throughout this paper, we adopt the natural unit $c=\hbar=1$.

\bfe{width=0.95\hsize}{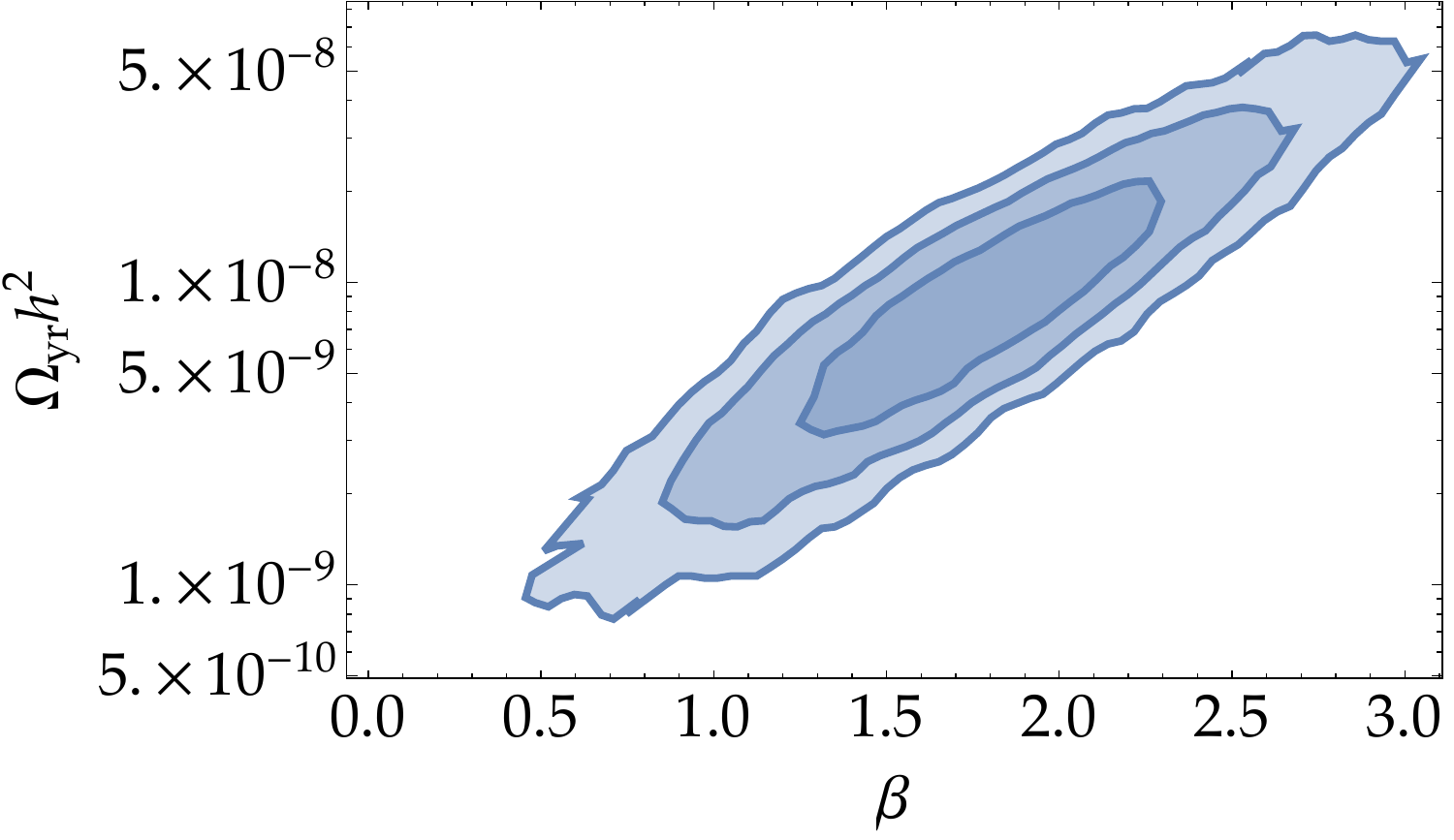}{The NANOGrav 15-year constraint ($1/2/3\sigma$) on the amplitude $\Omega_\yr h^2$ and the power $\beta$ for the power-law assumption $\Omega_\GW h^2(f)=\Omega_\yr h^2\pqty{f/f_{\si{yr^{-1}}}}^\beta$ (see Eq.~\eqref{eq: OGWh2} for the definition of the \ac{GW} density parameter $\Omega_\GW h^2$), translated from Fig.~11 of Ref.~\cite{NANOGrav:2023gor}.}{fig: NANOGrav const}

\section{Induced gravitational waves during the QCD phase transition}

We briefly review Ref.~\cite{Abe:2020sqb} for \ac{GW} induction during the \ac{QCD} phase transition.
Let us first specify the background dynamics.
The temperature dependence of the effective degrees of freedom for energy density, $g_*$, and entropy density, $g_{*s}$, of the \ac{QCD} plasma has been extensively studied both in analytic and numerical ways. Saikawa and Shirai unified these results in the form of the fitting function (see Appendix~C of Ref.~\cite{Saikawa:2018rcs}), which is plotted in the left panel of Fig.~\ref{fig: gs and wcs}.
These effective degrees of freedom are related to $w$ and $\cs^2$ by
\beae{\label{eq: w and cs2}
	&w(T)=\frac{4g_{*s}(T)}{3g_*(T)}-1, \\ 
	&\cs^2(T)=\frac{4(g_{*s}'(T)T+4g_{*s}(T))}{3(g_*'(T)T+4g_*(T))}-1,
}
which are shown in the right panel of Fig.~\ref{fig: gs and wcs}.
Once their temperature dependence is fixed, the time evolution of the temperature of the universe (and hence the evolution of all background parameters) can be calculated through the continuity equation, the Friedmann equation, and the definition of $g_*$:
\bege{
	\dv{\rho}{\eta}=-3(1+w)\calH\rho, \\ 
	3\Mpl^2\calH^2=a^2\rho \qc \rho(T)=\frac{\pi^2}{30}g_*(T)T^4.
}
$a$ is the scale factor, $\eta=\int a^{-1}\dd{t}$ is the conformal time, $\calH=\partial_\eta\ln a$ is the conformal Hubble parameter, and $\Mpl=1/\sqrt{8\pi G}$ is the reduced Planck mass.

Perturbations evolve along this background. The (Fourier-space) gravitational potential $\hat{\Phi}_\bfk(\eta)$ in the Newton gauge follows the Bardeen equation,
\bme{
	\hat{\Phi}_\bfk''(\eta)+3\calH(1+\cs^2)\hat{\Phi}_\bfk'(\eta) \\
	+\bqty{\cs^2k^2+3\calH^2(\cs^2-w)}\hat{\Phi}_\bfk(\eta)=0.
}
Here $\hat{\Phi}_\bfk(\eta)$ is decomposed into the transfer function $\Phi_k(\eta)$ and the primordial perturbation $\hat{\psi}_\bfk$ as $\hat{\Phi}_\bfk(\eta)=\Phi_k(\eta)\hat{\psi}_\bfk$. $\hat{\psi}_\bfk$ is related to the gauge-invariant curvature perturbation $\hat{\zeta}_\bfk$ by $\hat{\psi}_\bfk=-2\hat{\zeta}_\bfk/3$ and the initial condition of the transfer function is given by $\Phi_k(\eta)\to1$ and $\Phi_k'(\eta)\to0$ for $\eta\to0$.

The second-order effect of $\hat{\Phi}_\bfk(\eta)$ can source the linear tensor perturbation $\hat{h}_\bfk(\eta)$ through the equation,
\bae{
	\Lambda_\eta\pqty{a(\eta)\hat{h}_\bfk(\eta)}=4a(\eta)\hat{\calS}_\bfk(\eta),
}
where $\Lambda_\eta$ is the derivative operator
\bae{
	\Lambda_\eta=\partial_\eta^2+k^2-\frac{1-3w(\eta)}{2}\calH^2(\eta),
}
and $\hat{\calS}_\bfk(\eta)$ represents the source term
\begin{widetext}
\bae{
	\hat{\calS}_\bfk(\eta)=\int\frac{\dd[3]{\tilde{\bfk}}}{(2\pi)^3}e_{ij}(\bfk)\tilde{k}^i\tilde{k}^j\Biggl[2\hat{\Phi}_{\tilde{\bfk}}(\eta)\hat{\Phi}_{\bfk-\tilde{\bfk}}(\eta) 
	+\frac{4}{3(1+w(\eta))}\pqty{\hat{\Phi}_{\tilde{\bfk}}(\eta)+\frac{\hat{\Phi}_{\tilde{\bfk}}'(\eta)}{\calH(\eta)}}\pqty{\hat{\Phi}_{\bfk-\tilde{\bfk}}(\eta)+\frac{\hat{\Phi}_{\bfk-\tilde{\bfk}}'(\eta)}{\calH(\eta)}}\Biggr].
}
\end{widetext}
$e_{ij}(\bfk)$ is one polarization tensor.
This sourced equation can be solved in the Green function method as
\bae{
	\hat{h}_\bfk(\eta)=\frac{4}{a(\eta)}\int\dd{\tilde{\eta}}G_k(\eta,\tilde{\eta})\bqty{a(\tilde{\eta})\hat{\calS}_\bfk(\tilde{\eta})},
}
with the Green function $G_k(\eta,\tilde{\eta})$:
\bae{
	\Lambda_\eta G_k(\eta,\tilde{\eta})=\delta(\eta-\tilde{\eta}).
}
Practically, the Green function can be constructed by the two independent homogeneous solutions, $\Lambda_\eta g_{1k}(\eta)=\Lambda_\eta g_{2k}(\eta)=0$, as
\bae{
	G_k(\eta,\tilde{\eta})=\frac{g_{1k}(\eta)g_{2k}(\tilde{\eta})-g_{1k}(\tilde{\eta})g_{2k}(\eta)}{g_{1k}'(\tilde{\eta})g_{2k}(\tilde{\eta})-g_{1k}(\tilde{\eta})g_{2k}'(\tilde{\eta})}\Theta(\eta-\tilde{\eta}).
}
Eventually, the \ac{GW} power spectrum is given by
\begin{widetext}
\bae{\label{eq: calPh}
	\calP_h(k,\eta)=\frac{64}{81a^2(\eta)}\int_{\abs{k_1-k_2}\leq k\leq k_1+k_2}\dd{\ln k_1}\dd{\ln k_2}
	I^2(k,k_1,k_2,\eta)\frac{\pqty{k_1^2-(k^2-k_2^2+k_1^2)^2/(4k^2)}^2}{k_1k_2k^2}\calP_\zeta(k_1)\calP_\zeta(k_2),
}
where
\bae{
	I(k,k_1,k_2,\eta)=k^2\int^\eta_0\dd{\tilde{\eta}}a(\tilde{\eta})G_k(\eta,\tilde{\eta})\Biggl[2\Phi_{k_1}(\tilde{\eta})\Phi_{k_2}(\tilde{\eta}) 
	+\frac{4}{3(1+w(\tilde{\eta}))}\pqty{\Phi_{k_1}(\tilde{\eta})+\frac{\Phi_{k_1}'(\tilde{\eta})}{\calH(\tilde{\eta})}}\pqty{\Phi_{k_2}(\tilde{\eta})+\frac{\Phi_{k_2}'(\tilde{\eta})}{\calH(\tilde{\eta})}}\Biggr].
}
\end{widetext}
The \ac{GW} density parameter is well approximated by its oscillation average $\overline{\calP_h(k,\eta)}$ well after its horizon reentry as
\bae{
	\Omega_\GW(k,\eta)=\frac{\rho_\GW(\eta,k)}{3\Mpl^2H^2}=\frac{1}{24}\pqty{\frac{k}{\calH}}^2\overline{\calP_h(k,\eta)},
}
where $H=\calH/a$ is the ordinary Hubble parameter.
It is extended to the current time $\eta_0$ with the current radiation density parameter $\Omega_{r0}h^2=4.2\times10^{-5}$ as
\bae{\label{eq: OGWh2}
	\Omega_\GW(k,\eta_0)h^2=\Omega_{r0}h^2\pqty{\frac{a_\uc\calH_\uc}{a_\uf\calH_\uf}}^2\frac{1}{24}\pqty{\frac{k}{\calH_\uc}}^2\overline{\calP_h(k,\eta_\uc)},
}
where $h=H_0/(\SI{100}{km\, s^{-1} Mpc^{-1}})$ is the normalized Hubble constant, the subscripts `$\uc$' and `$\uf$' indicate the time when the \ac{GW} of interest becomes well subhorizon and the density parameter becomes almost constant (to which time we solve the induced \ac{GW}) and the time when all relevant phase transitions are completed and $g_*$ and $g_{*s}$ well asymptote to the current values (to which time we solve the background dynamics).

In this way, the induced \acp{GW} can be calculated on an arbitrary background.
In Fig.~\ref{fig: OmegaGW}, we show the example \ac{GW} spectrum normalized by the scalar amplitude squared $A_\zeta^2$ for the scale-invariant scalar perturbation $\calP_\zeta(k)=A_\zeta$.

\section{Gravitational wave signals}

\begin{figure*}
	\centering
	\begin{tabular}{c}
		\begin{minipage}{0.5\hsize}
			\centering
			\includegraphics[width=\hsize]{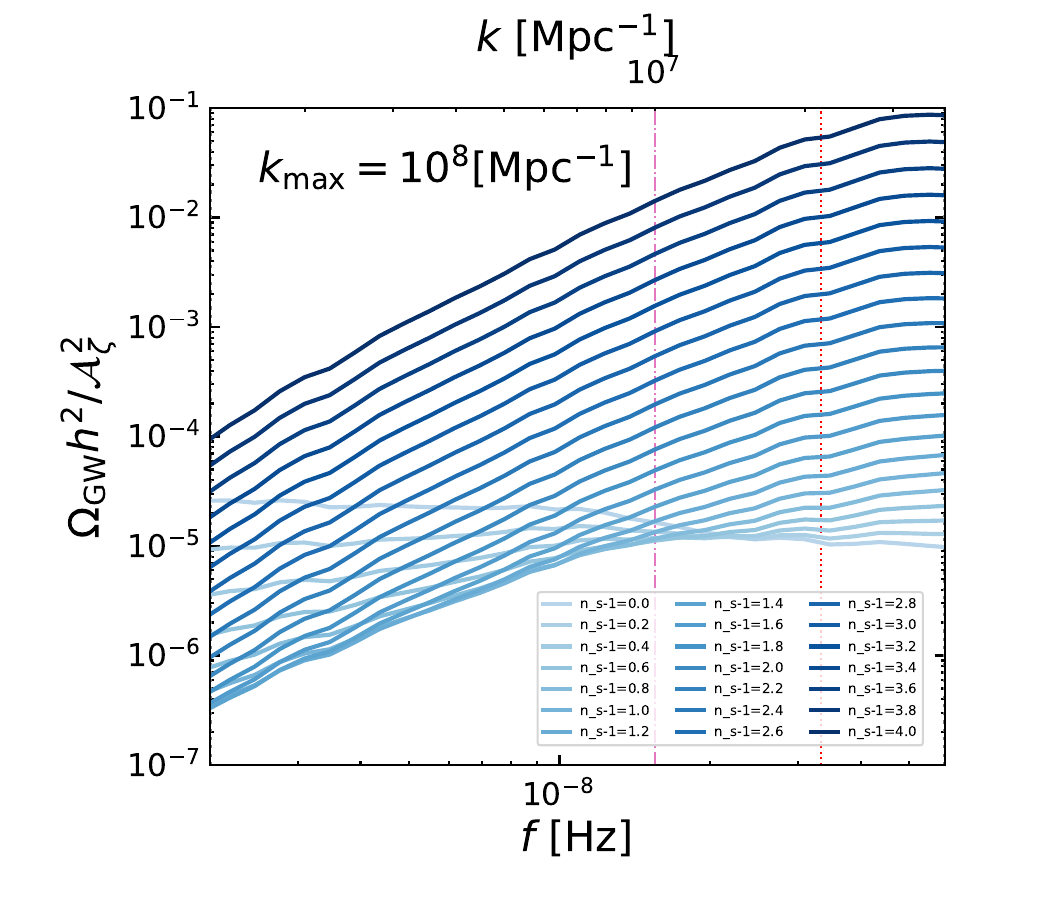}
		\end{minipage}
		\begin{minipage}{0.5\hsize}
			\centering
			\includegraphics[width=\hsize]{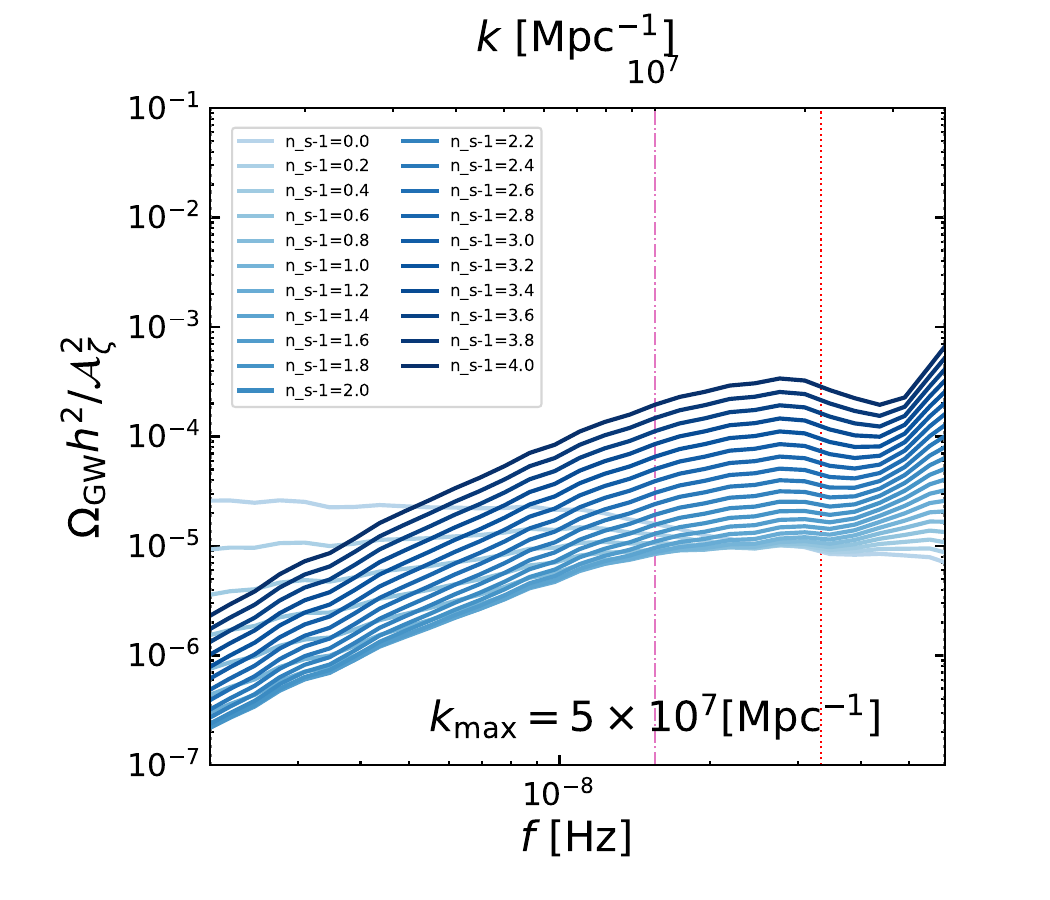}
		\end{minipage}
	\end{tabular}
	\caption{Resultant \ac{GW} spectra with $k_\mathrm{max}=10^8\,\si{Mpc^{-1}}$ (left) or $k_\umax=\SI{5e+7}{Mpc^{-1}}$ (right) for various values of the spectral index $\ns$ within the NANOGrav's sensitivity range $f=2\text{--}\SI{59}{nHz}$. The vertical red dotted line shows the pivot scale $f_{\si{yr^{-1}}}$ and the vertical pink dot-dashed line represents $k=10^7\,\si{Mpc^{-1}}$ below which the spectrum is fitted by the power-law function~\eqref{eq: OmegaGW fitting}.}
	\label{fig: various ns}
\end{figure*}

\begin{figure*}
	\centering
	\begin{tabular}{c}
		\begin{minipage}{0.5\hsize}
			\centering
			\includegraphics[width=\hsize]{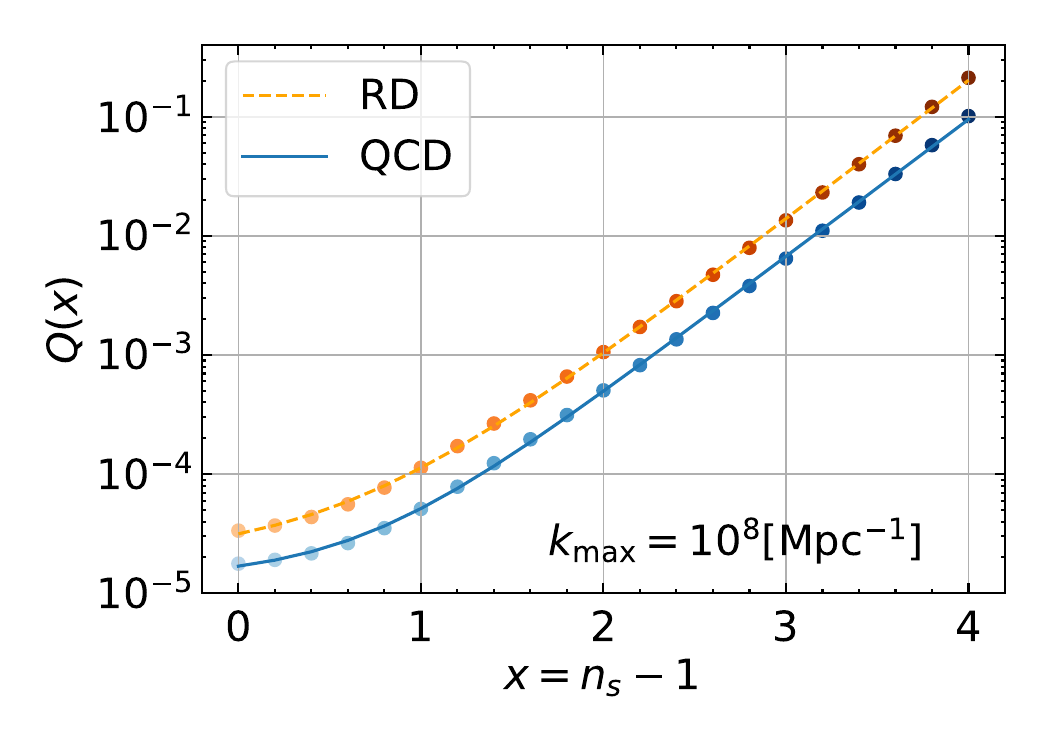}
		\end{minipage}
		\begin{minipage}{0.5\hsize}
			\centering
			\includegraphics[width=\hsize]{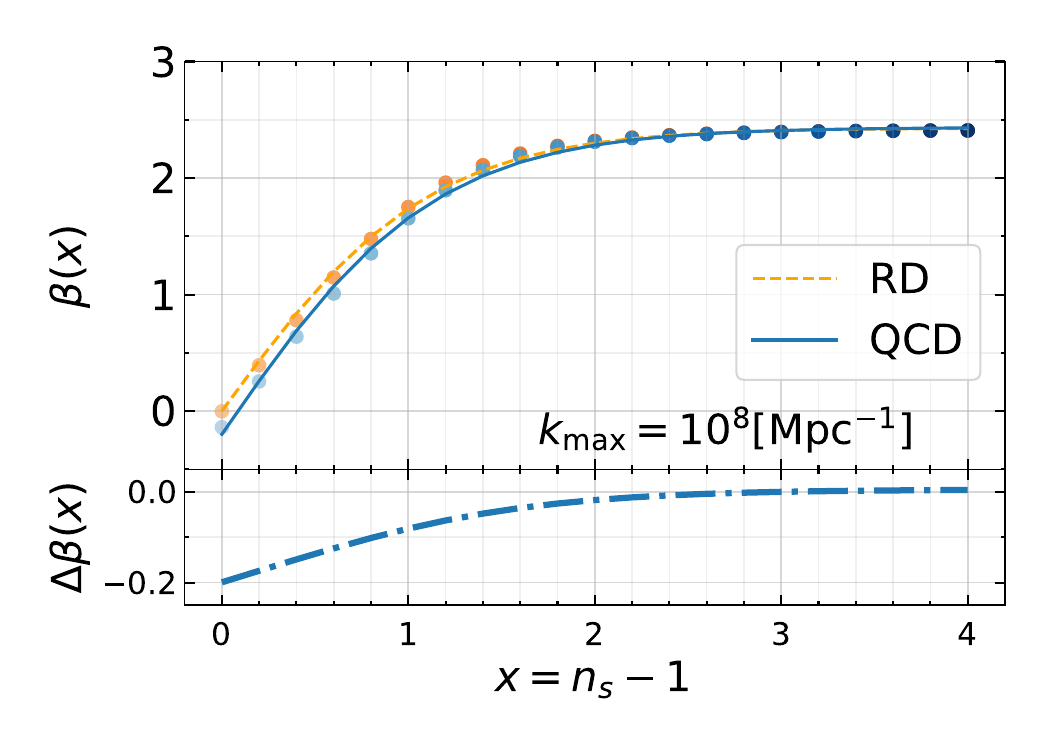}
		\end{minipage}
	\end{tabular}
	\caption{Fitting parameters $Q(x)$ (left), $\beta(x)$ (right top) and the difference $\Delta\beta(x)=\beta_\QCD(x)-\beta_\RD(x)$ (right bottom) in the power-law fitting~\eqref{eq: OmegaGW fitting} for numerical results (points) and their own fitting formulae~\eqref{eq: Q and beta fitting} (lines) for $k_\umax=10^8\,\si{Mpc^{-1}}$. The blue ones are the results with the \ac{QCD} effect while the orange ones are for the pure radiation-dominated universe as a comparison.
	}
	\label{fig: OGW and beta fit}
\end{figure*}

\begin{figure*}
	\centering
	\begin{tabular}{c}
		\begin{minipage}{0.5\hsize}
			\centering
			\includegraphics[width=\hsize]{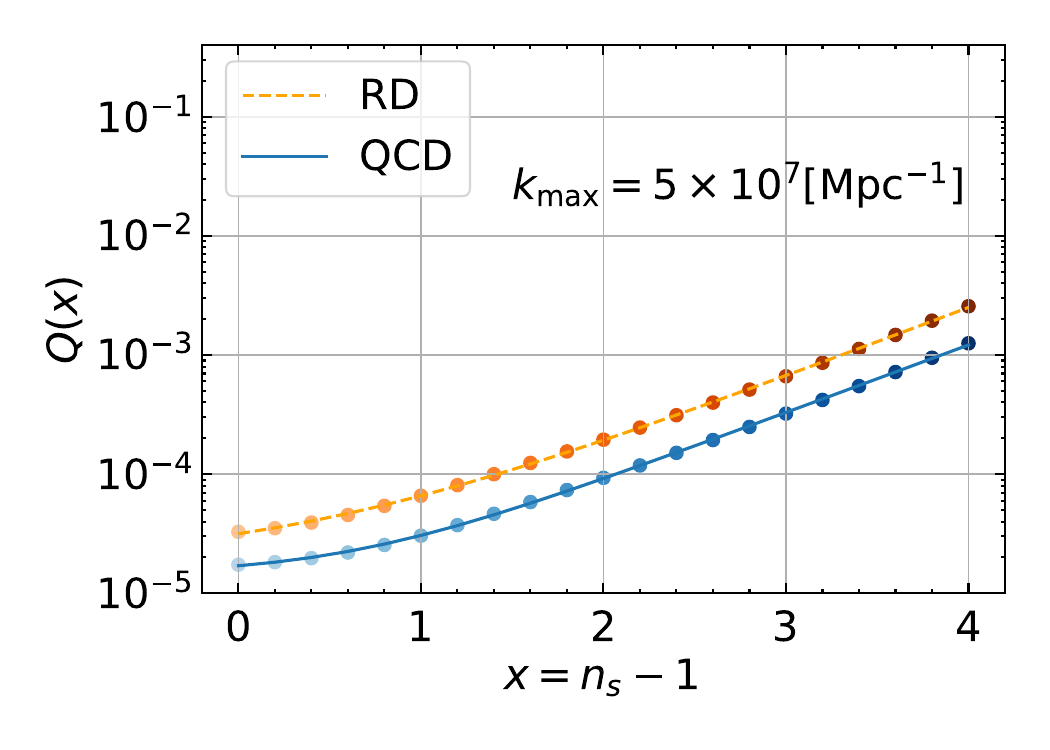}
		\end{minipage}
		\begin{minipage}{0.5\hsize}
			\centering
			\includegraphics[width=\hsize]{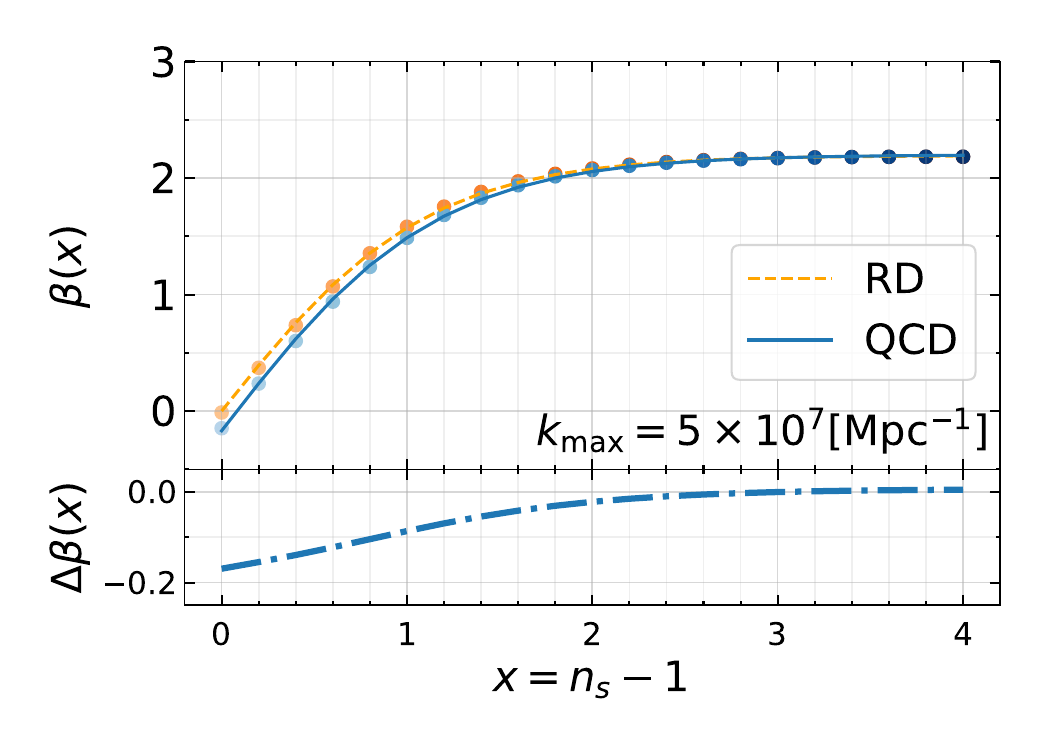}
		\end{minipage}
	\end{tabular}
	\caption{Same as Fig.~\ref{fig: OGW and beta fit} but with $k_{\mathrm{max}}=\SI{5e+7}{Mpc^{-1}}$. 
	}
	\label{fig: OGW and beta fit 5e7}
\end{figure*}

We then calculate the \ac{GW} spectrum, particularly in the NANOGrav's sensitivity range $f=2\text{--}\SI{59}{nHz}$ for the (compact) power-law-type power spectrum of the curvature perturbation:\footnote{Do not confuse this $\ns$ with the inferred value $\ns=0.965\pm0.004$ by the \ac{CMB} observation~\cite{Planck:2018vyg}. They correspond to different perturbation scales and are independent of each other in principle.}
\bae{\label{eq: power law power}
	\calP_\zeta(k)=A_\zeta\pqty{\frac{k}{k_{\si{yr^{-1}}}}}^{\ns-1}\Theta(k-k_\umin)\Theta(k_\umax-k),
}
where $k_{\si{yr^{-1}}}=2\pi\times\SI{1}{yr^{-1}}\simeq\SI{2e+7}{Mpc^{-1}}$ is the pivot scale and we restrict perturbations to a certain range $k_\umin\leq k\leq k_\umax$.
In particular, since NANOGrav's data favor the blue-tilted spectrum, the upper bound $k_\umax$ is practically necessary for the power spectrum not to exceed the unity and break the perturbativity. We take $k_\umax=10^8\,\si{Mpc^{-1}}$ as a fiducial value but the \ac{GW} spectrum can be sensitive to $k_\umax$ and we will also show the result for $k_\umax=\SI{5e+7}{Mpc^{-1}}$ as a comparison. $k_\umin$ is irrelevant in our setup and fixed to $k_\umin=10^5\,\si{Mpc^{-1}}$.
In Fig.~\ref{fig: various ns}, the resultant \ac{GW} spectra normalized by $A_\zeta^2$ are shown for several values of $\ns$.
Particularly in the low-frequency range to which pulsar timing array experiments are sensitive, they can be fitted by a power-law function
\bae{\label{eq: OmegaGW fitting}
	\Omega_\GW(f) h^2\approx Q(x)A_\zeta^2\pqty{\frac{f}{f_{\si{yr^{-1}}}}}^{\beta(x)},
}
with $x\equiv \ns-1$ and the fitting parameters $Q(x)$ and $\beta(x)$.
Figs.~\ref{fig: OGW and beta fit} and \ref{fig: OGW and beta fit 5e7} show fitting values of these parameters for each numerically-obtained \ac{GW} spectrum, where we only used the data of $k\leq10^7\,\si{Mpc^{-1}}$ for better fitting.
From this, one can further find the fitting formula for these fitting parameters themselves as
\beae{\label{eq: Q and beta fitting}
	&Q(x)\approx10^a (1+10^{(bx+c-a)/d})^d, \\
	&\beta(x)\approx e\tanh\pqty{fx}-g,
}
where the fitting parameters $a$, $b$, $c$, $d$, $e$, $f$, $g$ are summarized in Table~\ref{table: fitting params.}.
We note that while the power $\beta$ is given by an intuitive relation $\beta\sim2(\ns-1)$ in the nearly-scale-invariant case, $\beta$ shows a certain upper bound $\beta\lesssim3$ for larger $\ns$. In fact, it is known that for the broken-power-law primordial power spectrum, the low-frequency tail of \acp{GW} asymptotes to $\propto k^3$ if $\ns-1\geq3/2$~\cite{Atal:2021jyo,Domenech:2021ztg}. Therefore, $\beta$ cannot exceed three in our setup.
We have also confirmed that our numerical scheme is consistent with the analytic formulae derived in Refs.~\cite{Espinosa:2018eve,Kohri:2018awv}  in the radiation-dominated universe.\footnote{Some examples of the amplitude parameter $Q_\RD(\ns)$ (without the factor of $\Omega_{r0}h^2$) are shown in Table~1 of Ref.~\cite{Kohri:2018awv} but for the all integration range, $0<k_1,k_2<\infty$, in Eq.~\eqref{eq: calPh}. The same analysis is done by the European Pulsar Timing Array collaboration~\cite{Antoniadis:2023zhi}.}

\begin{table*}
	\caption{Fitting parameters in Eq.~\eqref{eq: Q and beta fitting} for the amplitude $Q(x)$ and the power $\beta(x)$ shown in Figs.~\ref{fig: OGW and beta fit} and \ref{fig: OGW and beta fit 5e7}.}
	\label{table: fitting params.}
	\begin{ruledtabular}
		\begin{tabular}{c|c|ccccccc}
			$k_\umax$ & background & $a$ & $b$ & $c$ & $d$ & $e$ & $f$ & $g$ \\
			\hline
			\multirow{2}{*}{$10^8\,\si{Mpc^{-1}}$} & QCD & $-4.854$ & $1.139$ & $-5.593$ & $1.068$ & $2.636$ & $0.8752$ & $0.1983$ \\
			& RD & $-4.684$ & $1.155$ & $-5.328$ & $1.420$ & $2.431$ & $0.8975$ & $0$ \\
			\hline
			\multirow{2}{*}{$\SI{5e+7}{Mpc^{-1}}$} & QCD & $-4.829$ & $0.5616$ & $-5.170$ & $0.6002$ & $2.370$ & $0.8648$ & $0.1686$ \\
			& RD & $-4.702$ & $0.5718$ & $-4.899$ & $0.9659$ & $2.194$ & $0.9020$ & $0$
		\end{tabular}
	\end{ruledtabular}
\end{table*}

\begin{figure*}
	\centering
	\begin{tabular}{c}
		\begin{minipage}{0.5\hsize}
			\centering
			\includegraphics[width=\hsize]{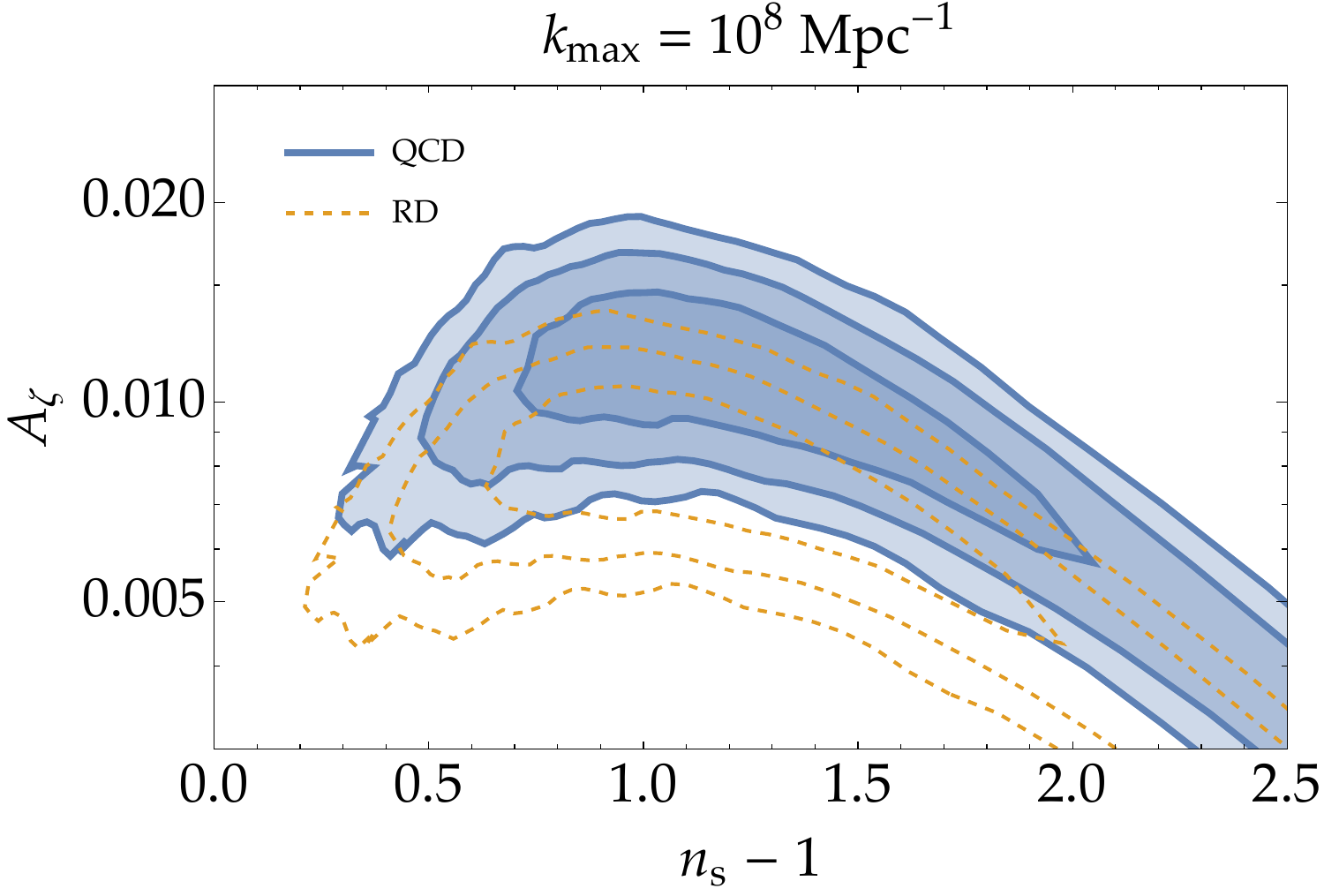}
		\end{minipage}
		\begin{minipage}{0.5\hsize}
			\centering
			\includegraphics[width=\hsize]{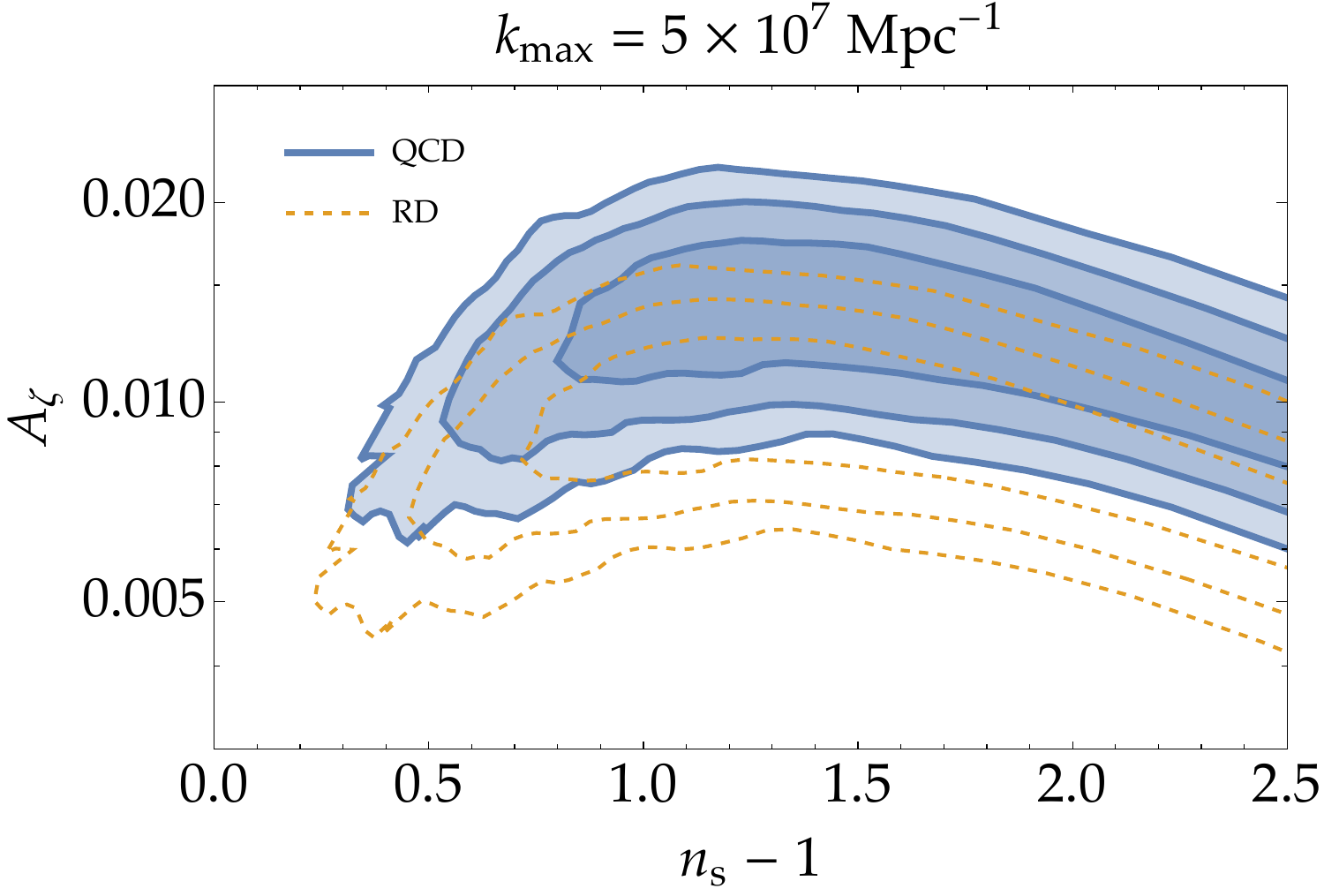}
		\end{minipage}
	\end{tabular}
	\caption{The inferred parameter region of the primordial power spectrum by the latest NANOGrav data (Fig.~11 of Ref.~\cite{NANOGrav:2023gor}) with (blue) and without (orange-dashed) the \ac{QCD} effect.}
	\label{fig: Az vs ns}
\end{figure*}

Making use of these formulae, the NANOGrav constraint on the \ac{GW} spectrum with the power-law assumption shown in Fig.~\ref{fig: NANOGrav const} can be interpreted as the constraint on the (compact) power-law-type primordial power spectrum as shown in Fig.~\ref{fig: Az vs ns}.
We also show the (mis)interpretation assuming the exact radiation background as a comparison. Additionally, we again stress the significance of imposing a practically required upper bound $k_{\mathrm{max}}$, on the power-law-type power spectrum. 
This constraint imposes an upper bound smaller than three on $\beta$. 
Consequently, the spectral index $n_{\mathrm{s}}-1$ is not constrained from above in contrast to Fig.~19 of Ref.~\cite{Antoniadis:2023zhi}, where the limitations $k_\umin$ and $k_\umax$ are not put.

\section{Summary and Discussion}

In this \emph{Letter}, we interpret the NANOGrav constraint on the \ac{GW} spectrum (Fig.~\ref{fig: NANOGrav const}) in terms of the parameters for the power spectrum of the primordial curvature perturbation~\eqref{eq: power law power}, including the \ac{QCD} phase-transition effect, as shown in Fig.~\ref{fig: Az vs ns}.
The constraint depends on the upper cutoff $k_\umax$ in Eq.~\eqref{eq: power law power} and the spectral index is not constrained from above, which is because the \ac{GW} spectrum with the power around or larger than three cannot be realized by the induced \acp{GW} from the broken-power-law primordial perturbation~\cite{Atal:2021jyo,Domenech:2021ztg}. 
However, one finds that anyway neglecting the \ac{QCD} effect would underestimate the amplitude of the primordial perturbation by about $25\%$. The spectral index could also be underestimated by up to $10\%$ when $0<n_{\mathrm{s}}-1<1$ if one neglects the \ac{QCD} effect.

We close the section by mentioning the implication of our result on the \ac{PBH} formation. 
The \ac{PBH} formation is also affected by the \ac{QCD} phase transition (see, e.g., Refs.~\cite{Jedamzik:1996mr,Byrnes:2018clq,Jedamzik:2020omx} and also Refs.~\cite{Escriva:2022bwe,Musco:2023dak} for detailed numerical studies on this effect) and the corresponding scenario is sometimes referred to as the \emph{thermal history} model~\cite{Carr:2019kxo,Clesse:2020ghq}.
Ref.~\cite{Carr:2023tpt} claims that this thermal history model is consistent with several observational ``positive evidence" of \acp{PBH} (see also the review article~\cite{Escriva:2022duf}).
Nevertheless, these works basically focused on the almost scale-invariant case $\ns\simeq0.96$ and cannot be directly applied to our spectrum $\ns-1\sim1$ inferred by the NANOGrav data.
Just regarding the amplitude, Ref.~\cite{Carr:2023tpt} supposes that the root-mean-square of the density contrast is $\sigma_\delta=0.0218$ on the solar-mass scale. $\sigma_\delta$ coarse-grained on a scale $R$ is related to the primordial power spectrum by (see, e.g., Ref.~\cite{Inomata:2016rbd})
\bae{
	\sigma_\delta^2=\frac{16}{81}\int\frac{\dd{k}}{k}W^2(kR)(kR)^4\calP_\zeta(k),
}
and for the compact power-law power spectrum $\calP_\zeta=A_\zeta(k/k_*)^{\ns-1}\Theta(k-k_\umin)\Theta(k_\umax-k)$ and the Gaussian window function $W(z)=\ee^{-z^2/2}$, it is simplified as
\bme{
	\sigma_{\delta,\PG}^2=\frac{8}{81}A_\zeta(k_*R)^{1-\ns} \\ 
    \times\bqty{\Gamma\pqty{\frac{\ns+3}{2},k_\umin^2R^2}-\Gamma\pqty{\frac{\ns+3}{2},k_\umax^2R^2}},
}
where $\Gamma(a,x)=\int^\infty_xt^{a-1}\ee^{-t}\dd{t}$ is the incomplete gamma function.
Making use of the mass-scale relation (see, e.g., Ref.~\cite{Tada:2019amh})
\bae{
	M(R)\simeq10^{20}\,\si{g}\pqty{\frac{g_*(R)}{106.75}}^{-1/6}\pqty{\frac{R}{\SI{6.4e-14}{Mpc}}}^2,
}
one finds that the inferred value, $A_\zeta\sim0.01$ and $\ns-1\sim1$ with $k_*=k_{\si{yr^{-1}}}$, $k_\umin=10^5\,\si{Mpc^{-1}}$, and $k_\umax=10^8\,\si{Mpc}^{-1}$ or $\SI{5e+7}{Mpc^{-1}}$, corresponds to $\sigma_{\delta,\PG}\sim0.015$ on the solar-mass scale, which could be consistent with Ref.~\cite{Carr:2023tpt}.
Detailed numerical studies for $\ns-1\sim1$ are anyway necessary.

\acknowledgments

Y.T. is supported by JSPS KAKENHI Grant
No.~JP21K13918.


\bibliography{main}
\end{document}